\documentclass{article}
\usepackage{dcase2020,amsmath,graphicx,url,times,booktabs, tabularx}
\usepackage{enumitem}
\usepackage{float}
\usepackage{multirow}
\usepackage{booktabs}
\usepackage[table,xcdraw]{xcolor}
\usepackage{soul}
\usepackage{mathtools}
\usepackage{textcomp} 
\usepackage[utf8]{inputenc}
\DeclareUnicodeCharacter{2212}{-}

\title{Anomalous Sound Detection using unsupervised and semi-supervised autoencoders and gammatone audio representation}

\name{Sergi Perez-Castanos$^{1}$,
        Javier Naranjo-Alcazar$^{1,2}$,
      Pedro Zuccarello$^{1}$, 
      Maximo Cobos$^{2}$,
      }

\address{$^1$ Visualfy, Benisan\'o, Spain \{sergi.perez, javier.naranjo, pedro.zuccarello\}@visualfy.com\\          
        $^2$ Universitat de Val\`encia, Burjassot, Spain, \{janal2\}@alumni.uv.es, \{maximo.cobos\}@uv.es\\
 }

\begin{document}

\ninept
\maketitle

\begin{sloppy}

\begin{abstract}
Anomalous sound detection (ASD) is, nowadays, one of the topical subjects in machine listening discipline. Unsupervised detection is attracting a lot of interest due to its immediate applicability in many fields. For example, related to industrial processes, the early detection of malfunctions or damage in machines can mean great savings and an improvement in the efficiency of industrial processes. This problem can be solved with an unsupervised ASD solution since industrial machines will not be damaged simply by having this audio data in the training stage. This paper proposes a novel framework based on convolutional autoencoders (both unsupervised and semi-supervised) and a Gammatone-based representation of the audio. The results obtained by these architectures substantially exceed the results presented as a baseline.
\end{abstract}


\section{Introduction}
\label{sec:intro}

\par Anomaly Sound Detection (ASD) has been receiving much interest from the scientific community in recent years \cite{kawaguchi2017can, koizumi2018unsupervised}. The early detection of anomalous events can mean a substantial improvement in systems that face  problems such as audio surveillance \cite{ntalampiras2011probabilistic, foggia2015audio} or predictive maintenance \cite{yamashita2006inspection, koizumi2017optimizing}.  This last case is related to the prediction and/or early detection of failures in industrial machinery and/or engines. This application is of particular interest since it could optimize and save a great amount of resources in industrial production chains.

\par The ASD problem can be separated into two categories: problems in which recordings of the anomalous events to be detected are available in the training phase and problems in which no anomalous events are available for training. The first type of problem is known as supervised-ASD  \cite{valenzise2007scream}, whereas the second is known as unsupervised-ASD \cite{patcha2007overview, chandola2009anomaly}. Supervised-ASD can be thought as a sound event detection (SED) problem, but with some peculiarities such as the duration and/or the nature of the sound event, like for example, a gunshot. On the other hand, in the unsupervised-ASD problem the objective is the detection of unknown or anomaly sound events without the system being aware of their existence, i.e., no anomalous events are available in the training data set. This would be the case in predictive maintenance of industrial machinery: it is unthinkable to damage, on purpose, machines of great economic cost just to obtain a set of audio samples. A good unsupervised-ASD system should be able to recognize the anomaly by training only with samples from non-anomalous, or normal, sound events.

\begin{figure*}[]
    \centering
    \centerline{\includegraphics[scale=0.5]{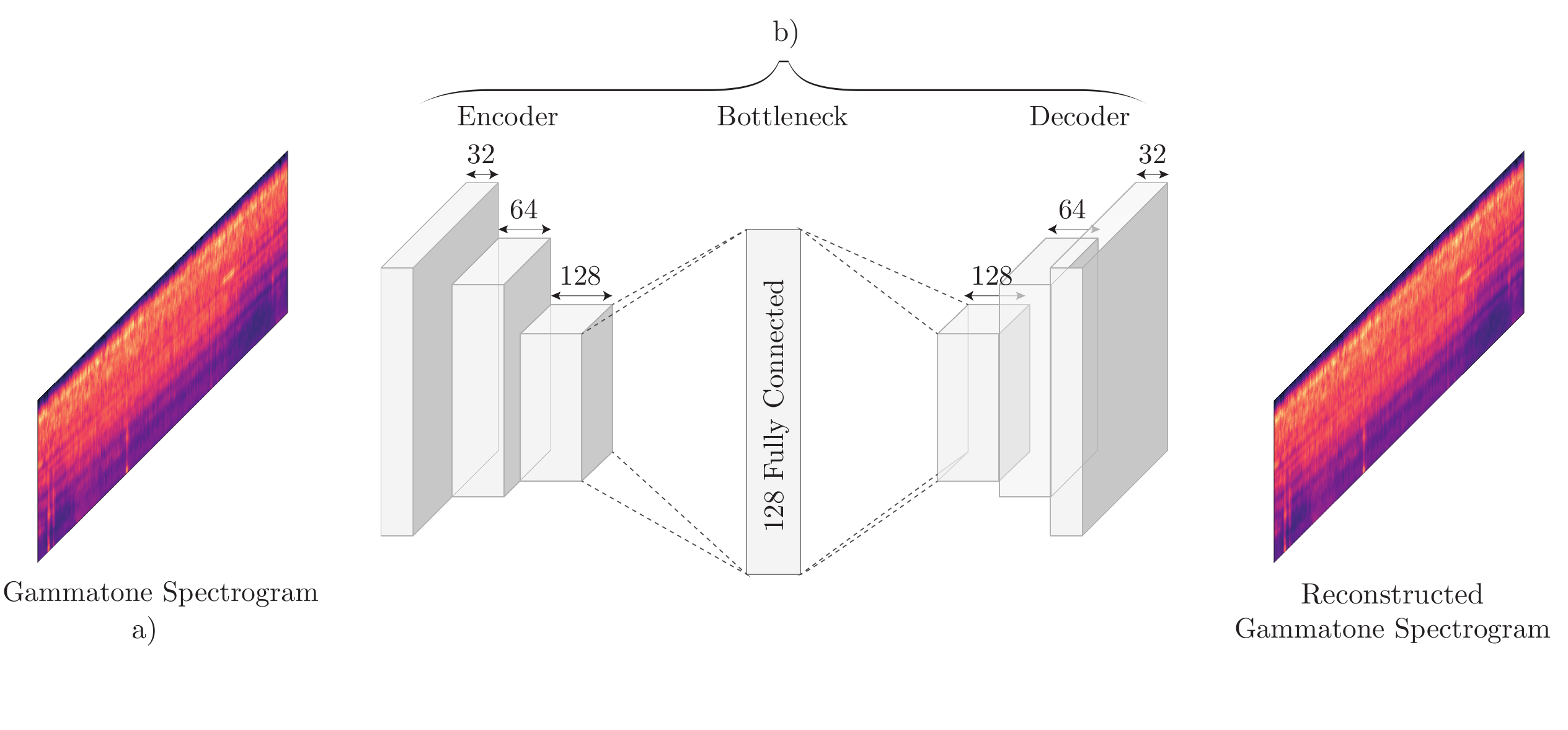}}
    \caption{Full framework for ASD based on a Convolutional Autoencoder. Step a) shows the chosen audio representation and step b) the designed autoencoder architecture. Each cube represents a ConvBlock. The numbers on top of each ConvBlock indicate the number of filters in each convolutional block.}
    \label{fig:autoencoder}
\end{figure*}

\par As it can be seen, this problem cannot be dealt as a classic classification problem like Acoustic Event Classification \cite{piczak2015environmental} or Audio tagging \cite{xu2017convolutional}. In this problem, there is a class, called \textit{unknown} or \textit{anomaly}, that must be recognized without the existence of positive samples of that class in the training set. In the case of engines or industrial machinery, the samples belonging to the \textit{anomaly} class, or  anomalous samples, are audio clips recorded when the machine is not working in the expected normal regime. The assumption is that this anomalous sounds show a different pattern than the ones produced with the machine working in normal regime. Therefore, if only one kind of training is available, a typical way of dealing with this kind of problem  would be an outlier-detection scheme, that is, calculating the deviation, or difference, between the normal samples and the anomalies, this value is known as anomaly score. If this value exceeds a certain threshold, the sample is considered anomalous.

\par The first approaches to the unsupervised-ASD problem were made using classic machine learning techniques such as Gaussian Mixture Models (GMMs) \cite{koizumi2017optimizing} or Support Vector Machines (SVMs) \cite{foggia2015audio}. In the last few years,  due to  the availability of larger amounts of data, Deep Learning techniques have become the state of the art in this field. As the main objective is to obtain a value, anomaly score, which provides us with information about the anomaly, the proposal of autoencoders seems to be a reasonable solution. Different architectures such as unsupervised autoencoders \cite{kawaguchi2017can, koizumi2018unsupervised, marchi2015novel, marchi2015non} have been proposed in the state of the art. These solutions often implement dense or recurrent layers instead of convolutionals. A different strategy may be the use of generative adversarial networks (GANs) \cite{goodfellow2014generative}. This type of network is composed of two modules: the generator and the discriminator. The first one is in charge of generating false samples and the second one of discerning if the sample is false or real. 


This work aims to propose a novel sound detection of anomalies based on a trained convolutional autoencoder with a 2D audio representation. The proposed scheme is applied to Task 2 of Detection and Classification of
Acoustic Scenes and Events (DCASE) 2020. The aim of this task is to identify malfunctioning states of a certain set of industrial machinery by analyzing its sound. As the sounds in abnormal functioning state are available, one of the proposed autoencoders is based on a semi-supervised architecture. Another architecture, based on an unsupervised classification scheme is also proposed and evaluated. In this last case, the malfunctioning information is not taken into account. The simplest approach would be to calculate an anomaly score per machine, that is, to train as many autoencoders as available machines. In this way, the autoencoder would be specialized to a certain type of machine. However, a more intersting and complex approach was chosen for this work: one single anomaly detector (in this case autoencoder) was trained for all the machines. This is, one single classifier is able to detect anomalies in the whole set of machines of the task.

\section{Task description and dataset}\label{sec:taskdataset}

Task 2 of the DCASE 2020 edition is the first to introduce the issue of ASD into this challenge. The objective of this task is to perform an \textit{Anomalous Sound Detection System} ($\mathcal{A}_{\theta}$) that is able to identify anomalies in different audio samples produced by industrial machines. This problem has nothing to do with a classification problem between normal and anomalous classes because only normal samples are available when training the system. Therefore, the anomalous class is unknown to the $\mathcal{A}_{\theta}$. This fact has been the main difference between this task and others presented in the DCASE that also presented the problem of anomalous detection but in a supervised way \cite{DCASE2017challenge}.

The dataset used to train and evaluate models is the one presented in the task, focused on ASD. It consists in subsets of ToyADMOS \cite{koizumi2019toyadmos} and MIMII \cite{purohit2019mimii} datasets. From the first one, car and conveyor classes are combined with valve, pump, fan and slide rail classes from the second one. In this context, a class corresponds to a machine type.

The audios have been recorded with a sampling rate of 16 kHz. Each class is divided into 2 groups: normal sounds and anomalous sounds, that are those sounds that belongs to damaged machines. Audio clips are divided into 2 folds: train and test. In train fold, only normal sounds are taken into account while both types are included in test fold. In subsection~\ref{subsec:training} explains the process of how the samples used during the training stage are divided into two subgroups: training and validation. The validation samples are used to update the callbacks and to choose the model that will best generalize the test set. 

\section{Proposed method}
\label{sec:method}

\par The proposed method is constituted by two steps: a 2D audio representation and a convolutional autoencoder with a bottleneck layer that acts as a divider between the encoder and the decoder. It is important to emphasize that a single autoencoder is trained for all available machines. As mentioned in the task description, this solution is much more challenging than proposing one autoencoder per machine type.

\subsection{Audio representation}\label{subsec:audio}

The 2D audio representation used in this framework is based on Gammatone filters \cite{tabibi2017investigating}. This filter bank has shown promising results in the task of audio classification, surpassing the representation based on Mel filters \cite{zhang2018deep}, proposed, for example, in the MIMII dataset baseline \cite{purohit2019mimii}. Temporal bins are calculated with a window size of 40~ms and an overlap of 50\%. The number of filters or frequency bins is set to 64. Once the representation is obtained, the logarithm is calculated and a normalization of zero mean and standard deviation of one is performed for each frequency bin with all available data. Therefore, the representation has a size of $64 \times T \times 1$, where $T$ corresponds to the temporal bins according to the duration of the audio.

\subsection{Autoencoder architecture}\label{subsec:autoencoder}

The autoencoder is made up of convolutional layers and a dense layer acting as bottleneck. As can be seen in Figure~\ref{fig:autoencoder}, the encoder and decoder have a symmetric architecture. As can be recognized, each one is composed of 3 convolutional blocks (ConvBlocks). Each convolutional block is actually composed of 7 layers. The convolutional layer with $3 \times 3$ kernel size, the batch normalization (BN) layer and the activation layer, in this case Rectified Linear Unit (ReLU). This set of 3 layers is repeated twice and followed by a pooling layer. In the case of the decoder, the pooling layer is replaced by an upsampling layer. ConvBlocks can be found in Figure~\ref{fig:convblocks}. The bottleneck layer corresponds to a Dense layer of 128 neurons with linear activation. This layer is the least dimensional representation that the encoder makes of the input signal and from which the decoder must be able to reconstruct to obtain the same input signal. Unlike the encoder, the decoder has an extra convolutional layer with 1 filter and linear activation that is responsible for reconstructing the representation of the input.

The architecture explained previously corresponds to an unsupervised autoencoder, that is, its only purpose is to reconstruct the input without taking into account extra information such as the class of the input signal. Therefore, the cost function to be optimized in this architecture is the mean squared error (MSE).

\begin{figure}[]
    \centering
    \centerline{\includegraphics[width=0.8\columnwidth]{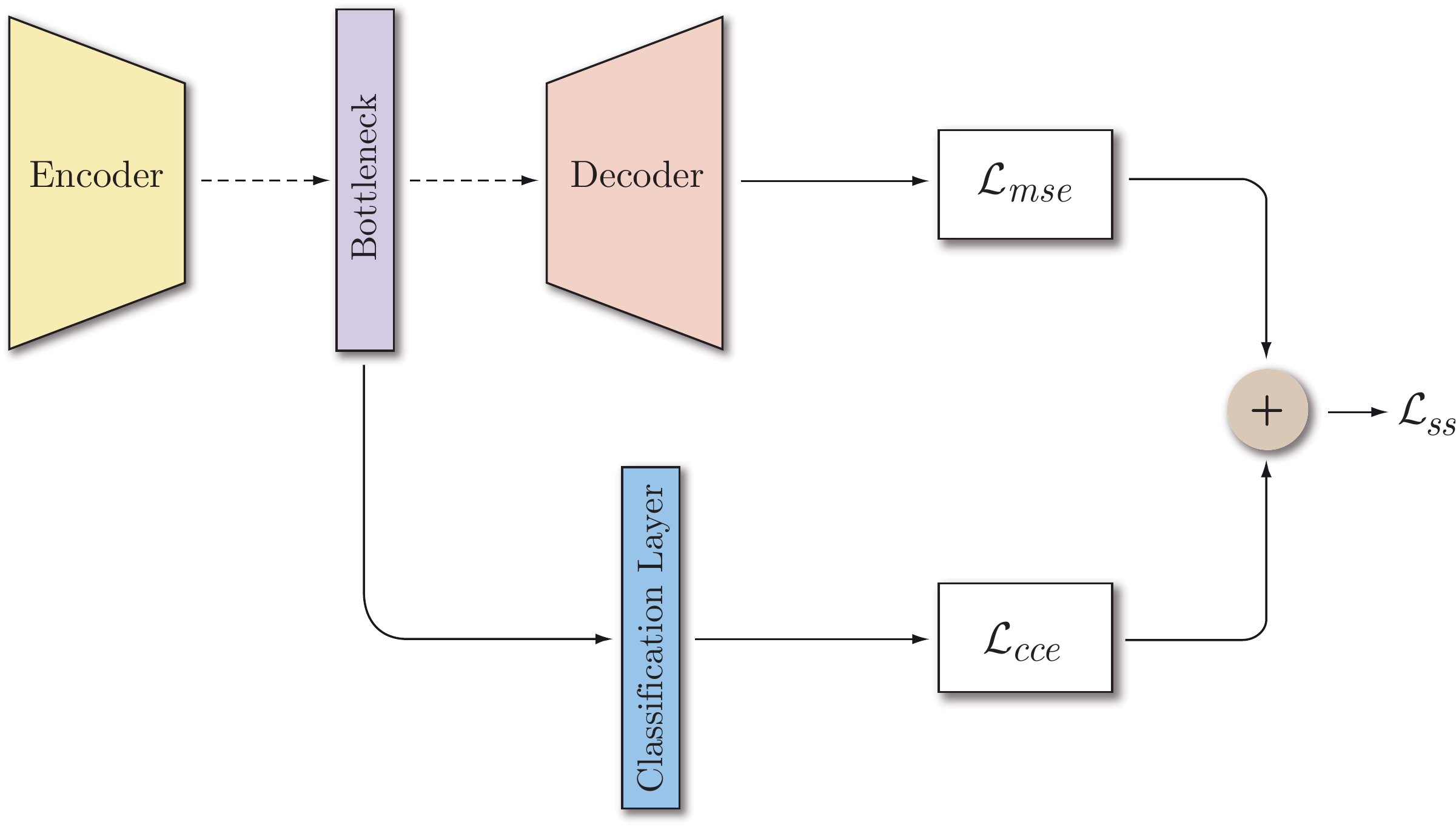}}
    \caption{Semi-supervised autoencoder architecture}
    \label{fig:semisupervised}
\end{figure}

\begin{table*}[]
\centering
\begin{tabular}{ccccccc}
\toprule
Framework       & \multicolumn{6}{c}{AUC}          \\ \cmidrule(lr){2-7}
                & ToyCar & ToyConveyor & fan & pump & slider & valve  \\ \midrule 
B        &  78.77$\pm$1.03      &   72.53$\pm$0.67          &  65.83$\pm$0.53   & 72.89$\pm$0.70     &   84.76$\pm$0.29     &  66.28$\pm$0.49         \\ \midrule
U  & \textbf{95.67}  &    \textbf{96.63}    &      79.87       &  81.51   &  80.86    &   82.85              \\ \midrule
U FD & 91.12  &    93.36    &      \textbf{80.40}       &  \textbf{82.61}   &  \textbf{81.16}    &   \textbf{83.19}              \\ \midrule
SS-0.7-0.3 &   87.27     &      90.35       &   78.63  &   80.33   &    78.94    &     80.94     \\ \midrule
SS-0.5-0.5 &   73.16    &     80.82        &  70.82   &  71.84    &   70.53    &    71.77 \\ \midrule
SS-0.3-0.7 &   63.82    &     74.65        &  63.41   &  64.09    &   62.15     &    64.18     \\
\bottomrule
\end{tabular}
\caption{AUC (\%) obtained by the proposed frameworks compared to the baseline. B means baseline, U stands for unsupervised autoencoder, SS represents semi-supervised autoencoder and FD denotes that full dataset, composed of 1st and 2nd releases, was used in training stage. SS values are followed by the values of $\alpha$ and $\beta$ (see Eq.~\ref{eq:lss}).}
\label{tab:results}
\end{table*}

By having the information of what type of machine is associated with each audio, the autoencoder can be modified to take this into account. This would correspond to a  semi-supervised scheme. The labeled supervised information is the type of machine; the labels do not indicate whether the audio clip corresponds to normal or abnormal functioning state of the machine \cite{kawachi2018complementary}. In the semi-supervised case, a dense classification layer with a number of units equal to the number of classes (machines) in the dataset is added. This layer takes the bottleneck as input, which, as previously explained, is where the the highest degree of compression is achieved. Figure~\ref{fig:semisupervised} depicts the proposed scheme. In the semi-supervised case, the cost function is affected and the classification error is now taken into account by means of the categorical crossentropy loss (CCE):

\begin{equation}
    \mathcal{L}_{ss} = \alpha \mathcal{L}_{mse} + \beta \mathcal{L}_{cce}
\label{eq:lss}
\end{equation}

where $\mathcal{L}_{ss}$ is the loss value of the semi-supervised architecture, $\mathcal{L}_{mse}$ corresponds to the mean squared error and $\mathcal{L}_{cce}$ represents the categorical crossentropy loss. $\alpha$ and $\beta$ are weighting factors such that $\alpha + \beta = 1$.

\begin{figure}[]
    \centering
    \centerline{\includegraphics[width=0.68\columnwidth]{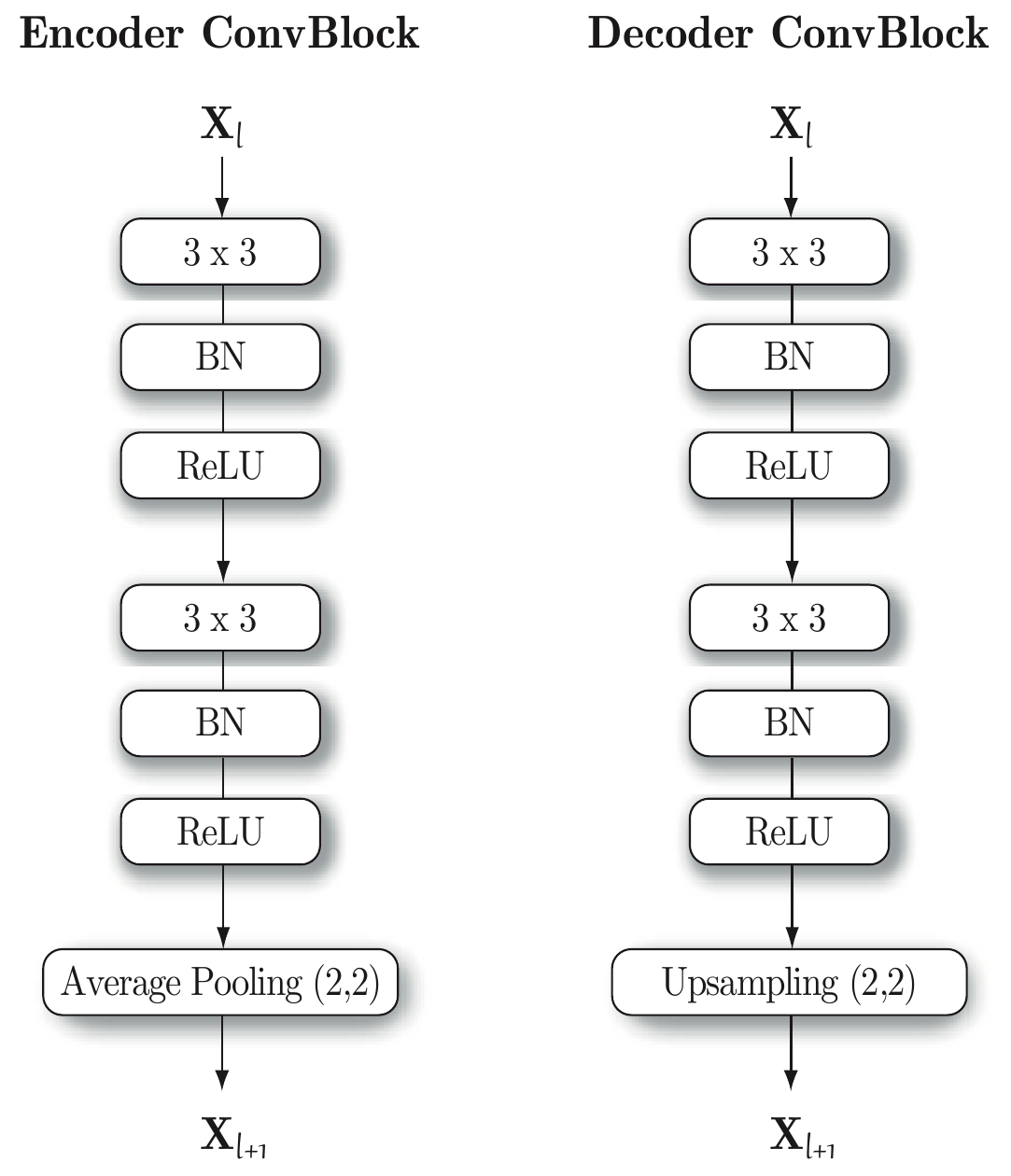}}
    \caption{Design of the convolutional blocks used in both the encoder and the decoder}
    \label{fig:convblocks}
\end{figure}

\section{Experimental details}\label{sec:exp_details}




\subsection{Training procedure}\label{subsec:training}

The training process for the two autoencoder architectures has been the same. The batch size is set to 32. The system is trained for a maximum of 500 epochs. If the validation loss does not improve by 20 epochs, the learning rate decreases by a factor of 0.75. If this metric does not improve by 50 epochs, the training is terminated. The optimizer used was Adam \cite{kingma2014adam}. The validation set corresponds to 10\% of the available samples in training.

\section{Results}\label{sec:results}


\subsection{Baseline system}\label{subsec:baseline}
The baseline system of Task 2 is a semi-supervised scheme similar to the one shown in Figure~\ref{fig:semisupervised}. The audio is represented with a 128-filter Mel filter-bank. The window size is 64~ms with a 50\% overlap. However, the input of the autoencoder is a 640-position vector. This is because 5 time-frames are concatenated: the frame of time \emph{t} is concatenated with the frames of times \emph{t-2},  \emph{t-1},  \emph{t+1} and  \emph{t+2}. Both the encoder and the decoder are composed of 4 Dense layers of 128 units followed by a batch normalization layer and ReLU activation.  The bottleneck is composed of 8 units and, unlike the autoencoder proposed in this work, it is also followed by a normalization and ReLU activation layer. The last layer of the decoder, the fifth in this case, is composed of 640 units as well as the input size.

\begin{table*}[]
\centering
\begin{tabular}{ccccccc}
\toprule
Framework       & \multicolumn{6}{c}{pAUC}          \\ \cmidrule(lr){2-7}
                & ToyCar & ToyConveyor & fan & pump & slider & valve  \\ \midrule 
B        &  67.58$\pm$1.04      &   60.43$\pm$0.74          &  52.45$\pm$0.21   & 59.99$\pm$0.77     &   66.53$\pm$0.62     &  50.98$\pm$0.15        \\ \midrule
U    &    \textbf{87.14}    &      \textbf{90.45}       &  70.78   &  70.99    &    \textbf{70.69}    &    71.62        \\ \midrule
U FD    &    73.41    &      80.32       &  \textbf{72.56}   &  \textbf{72.23}    &    69.94    &    \textbf{72.34}        \\ \midrule
SS-0.7-0.3 &   74.21    &     81.50        &  71.26   &  70.94    &   70.08     &    70.83      \\
\midrule
SS-0.5-0.5 &   60.42    &     71.63        &  60.32   &  58.88    &   58.51     &    58.70      \\ \midrule
SS-0.3-0.7 &   55.58    &     68.18        &  57.33   &  55.67    &   55.07     &    55.39       \\\bottomrule
\end{tabular}
\caption{pAUC (\%) obtained with the proposed frameworks compared to the baseline. Notation is the same as in Table~\ref{tab:results}}
\label{tab:results_pauc}
\end{table*}

\subsection{Metrics}\label{subsec:metrics}

The metrics used to evaluate the systems are the area under the curve (AUC) under the receiver operating characteristic (ROC) and the partial-AUC (pAUC). Both metrics are calculated from a portion of the ROC curve over a pre-specified range of interest. In this work, pAUC is computed as the AUC over a low false-positive-rate (FPR) range $[0,p]$. Therefore, metrics are expressed as:

\begin{equation}
    {\rm AUC} = \frac{1}{N_{-}N_{+}} \sum_{i=1}^{N_{-}} \sum_{j=1}^{N_{+}} \mathcal{H} (\mathcal{A}_{\theta} (x_{j}^{+}) - \mathcal{A}_{\theta} (x_{i}^{-})),
\end{equation}

\begin{equation}
    {\rm pAUC} = \frac{1}{\lfloor p N_{-} \rfloor N_{+}} \sum_{i=1}^{\lfloor p N_{-} \rfloor} \sum_{j=1}^{N_{+}} \mathcal{H} (\mathcal{A}_{\theta} (x_{j}^{+}) - \mathcal{A}_{\theta} (x_{i}^{-})),
\end{equation}

where $\lfloor.\rfloor$ denotes the flooring function. $\mathcal{H} (x)$ is a function that return 1 if $x > 0$ and 0, otherwise. $\{x_{i}^{−}\}_{i=1}^{N_{−}}$ and $\{x_{j}^{+}\}_{j=1}^{N_{+}}$ are the normal and anomalous test samples, respectively. Samples have been sorted so that the anomaly scores are in descending order. Therefore, $N_{−}$ and $N_{+}$ are the number of samples of the normal and anomalous category respectively. $\mathcal{A}_{\theta} (x)$ denotes the anomaly score of a given audio and according to the previous formula, this measure among the the normal test samples are used as the threshold. The use of pAUC is based on applications requirements. An ASD systems that alters falsely many times cannot be trusted. Therefore, it is important to increase the true-positive-rate under low FPR conditions. In this work, $p$ is set to $0.1$. This metric emphasizes the trade-off that must be made when implementing this type of system.

\subsection{Analysis}\label{subsec:analysis}

Results obtained in this task are shown in Tables~\ref{tab:results} and \ref{tab:results_pauc}. As it can be appreciated, almost all the proposed frameworks exceed the results presented as a baseline \cite{koizumi2020description} except the class \textit{slider} which cannot be improved by any proposed system in the AUC metric. The architecture that shows a better result is the unsupervised one. However, some machines show a better result when the training set corresponds only to the portion released in the first release.

As it can be observed, the improvement is substantial in all machines obtaining the lowest improvement in the \textit{pump} class of about 10 percentage points. On the other hand, \textit{ToyConveyor} is the machine that has been most improved with about 24 more percentage points compared to the baseline. As far as the slider machine is concerned, a decrease of about 4 percentage points is obtained.

As for the semi-supervised architecture, the grid search performed with the $\alpha$ and $\beta$ values shows that the more weight is given to the classification error the worse are the obtained results. The worst system is the one in which $\beta = 0.7$. As it can be seen, it was decided not to train these systems with the whole training set as it was not going to present any significant improvement.  Therefore, it can be deduced that such extra information from the machine's label adds noise to the bottleneck leading to a worse reconstruction by the decoder. The purely unsupervised system shows the best behaviour for this problem.




\section{Conclusion}
\label{sec:con}

The state of the art in the field of Anomalous Sound Detection has shown the great potential that solutions based on autoencoders have for mitigating the problems related to this task. In previous anomaly detection works, different architectures have been proposed, such as variational autoencoders \cite{kawachi2018complementary}, however, approaches with  autoencoders based on convolutional layers are not so common in the literature. Therefore, this paper shows the potential of this type of layers to extract relevant information from the audio in order to obtain the necessary anomaly score to discern whether the sample is anomalous or not. In addition, it is also studied how a semi-supervised architecture behaves in this kind of problems. Regarding the audio representation, the choice was made to use the Gammatone representation instead of using Mel filter-banks, or even instead of converting the audio into a one dimensional vector as it is proposed in state of the art solutions \cite{koizumi2020description}.

\section{ACKNOWLEDGMENT}
\label{sec:ack}

\par The participation of Javier Naranjo-Alcazar and Dr. Pedro Zuccarello in this work is partially supported by Torres Quevedo fellowships DIN2018-009982 and PTQ-17-09106 respectively from the Spanish Ministry of Science, Innovation and Universities. The work of Maximo Cobos was supported in partby the European Regional Development Fund (ERDF), and in part by the Spanish Ministry of Science, Innovation, and Universities under Grant RTI2018-097045-B-C21.

\bibliographystyle{IEEEtran}
\bibliography{refs}

\end{sloppy}
\end{document}